\documentclass[aps,prd,preprint,superscriptaddress,showpacs,nofootinbib]{revtex4}
\usepackage{slashed}
\usepackage{epsfig,latexsym,cancel,amssymb,amsmath,verbatim,mathrsfs}
\usepackage{color}
\usepackage{graphicx}

\def\ra{\rightarrow}
\def\L{\left(}
\def\R{\right)}

\def\Ld{\Lambda}
\def\ld{\lambda}
\def\f{\frac}
\newcommand{\be}{\begin{equation}}
\newcommand{\ee}{\end{equation}}
\newcommand{\bea}{\begin{eqnarray}}
\newcommand{\eea}{\end{eqnarray}}
\newcommand{\ba}{\begin{array}}
\newcommand{\ea}{\end{array}}

\long\def\symbolfootnote[#1]#2{\begingroup%
\def\thefootnote{\fnsymbol{footnote}}\footnote[#1]{#2}\endgroup}

\newcommand{\beq}{\begin{equation}}
\newcommand{\eeq}{\end{equation}}

\begin{document}

\title{Upgrading Sterile Neutrino Dark Matter to FI$m$P \\ Using Scale Invariance}

\date{\today}

\author{Zhaofeng Kang}
\email[E-mail: ]{zhaofengkang@gmail.com}
\affiliation{School of Physics, Korea Institute for Advanced Study,
Seoul 130-722, Korea}

\begin{abstract}

In this article we propose a class of extremely light feebly interacting massive particle, FI$m$Ps. They are combination of feebly interacting massive particle with scale invariance, by which DM stability, mass origin and relic density are inherently related. In the scale invariant version of the Standard Model (SM) with three  right-handed neutrinos ($\nu$SISM), the lightest $N_1$ realizes the FI$m$P scenario. In this example scalar singlets, which are intrinsic to the $\nu$SISM, generate mass and relic density for this FI$m$P simultaneously. Moreover, they are badly needed for electroweak symmetry spontaneously breaking. Interestingly, a 7.1 keV $N_1$ with correct relic density, that can explain the recent 3.55 keV $X-$ray line, lies in the bulk parameter space of our model.

\end{abstract}

\maketitle

\section {introduction and motivation}

The existence of dark matter (DM) is commonly believed, nevertheless, its particle properties are still in the dark. The hypothesis that DM is a weakly interacting massive particle (WIMP) prevails, mainly due to the so-called WIMP miracle, which says that if DM annihilates involving weak scale mass and strength, it will obtain the correct order of magnitude for relic density $\Omega_{\rm DM} h^2\sim 0.1$. Despite of a lot of experimental efforts (including (in)direct DM detections and also collider searches), people still have not detected any trace of DM with confirmation. These null results~\cite{LUX} begin to challenge the WIMP paradigm.

Now it is at the right time to reexamine the theoretical basis of WIMP DM and explore other DM frameworks, or more concretely, models. We would like to follow the ensuing basic questions about DM. 
\begin{itemize}
\item First, why is it there? A DM candidate that is predicted rather than introduced would be more attractive. A good case in point is the lightest sparticle in SUSY and axion from the Peccei-Quinn models aiming at solving the strong CP problem. Another example is the core of this paper, the lightest right-handed neutrino (RHN) $N_1$, or sterile neutrino~\cite{osci}. Originally RHNs are introduced to explain the nonzero neutrino masses, but they are neutral and thus potential to provide a DM candidate.

\item Second, why is it stable (or at least sufficiently long-lived at cosmic time scale)? Again, we advocate a mechanism that naturally guarantees DM stability rather than imposing a protective symmetry like $Z_2$ by hand. For instance, such a $Z_2$ may be an accidental symmetry due to the field content along with the space-time and gauge symmetries of the model, see an example based on scale invariance (SI)~\cite{Guo:2014bha,SIBL1}~\footnote{ (Under some reasonable assumptions) it concludes that the only viable accidental DM (aDM) must be a singlet scalar. But this paper is restricted to WIMP. Beyond it, a singlet fermion $N$ is also viable, provided that its coupling to the visible sector is sufficiently weak, namely $y_N\bar \ell \Phi N$ with $y_N\ll1$. For $N$ introduced in this way is naturally identified with the lightest RHN.}, a symmetry which may address the hierarchy problem~\cite{Bardeen,SI:old,SI:hidden,SI:boson,SI:DM,SI:strong}. Sterile neutrino DM offers another line for stability: lightness along with feebly interactions. No exact symmetry, by hand or accidental, is invoked here and thus DM is expected to decay, say, into the very light SM species like neutrino and photons. But the decay rate is greatly suppressed by powers of DM mass and couplings, so DM can be sufficiently long-lived.

\item Last but not the least, why is the DM relic density $\Omega_{\rm DM} h^2$ around 0.1? The WIMP scenario presents the WIMP miracle~\footnote{In practice, a lot of WIMP dark matters studied recently are far away from this miracle.}. We find that similar numerical coincidence may also arise in the framework of feebly interacting massive particle (FIMP)~\cite{freezein}. Moreover, it is tied to the mass origin and stability of DM.

 \end{itemize}
 
Neglect the first question for the time being. The answer to the second and third questions may point to a  framework which combines FIMP with SI. The FIMP DM $X$ (e.g., a Majorana fermion like $N_1$) feebly interacts with other fields, including the Higgs-like field ${\cal S}$ that generates its mass via the term ${\cal S}X^2$, so it will be very light as long as the vacuum expectation value (VEV) of ${\cal S}$ is not hierarchically above TeV. Lightness plus feeble interactions are appropriate to give accidentally long-lived DM. Additionally, that mass source term can simultaneously freeze-in FIMP with correct relic density. Hereafter that kind of FIMP is dubbed as FI$m$P with ``$m$" indicating its lightness due to SI. Needless to say, FI$m$P can not be seen by the current DM detectors, except for these sensitive to $X-$ray. As a mention, with regard to experimental tests, FI$m$P, which tends to produce $X-$ray, is more interesting than FIMP which usually stays in the dark.

The FI$m$P example $N_1$ from the $\nu$SISM is of special interest, because it gives answers to all those three questions. Interestingly, there are some suggestive hints which favor such DM in the $X-$ray line at energy 3.55 keV, that was reported recently and can be explained by the 7.1 keV $N_1$ with active-sterile neutrino mixing angle $\simeq 0.8\times10^{-5}$~\cite{Boyarsky:2014jta}. Despite of the difficulty in the conventional ways, such a $N_1$ can easily get correct relic density via freeze-in~\cite{Frigerio:2014ifa,Adulpravitchai:2014xna,Merle:2014xpa}. In this paper we will concentrate on the $N_1$ example, and another interesting example based on scalar FI$m$P is considered in Appendix.~\ref{SFImP}.

In the late stage of this paper, we found that the idea of introducing a scalar singlet to freeze-in $N_1$ has already been explored by several groups~\cite{inflaton,Kusenko:2006rh,Merle:2013wta}. Even then, there are several obvious difference between our paper and theirs. First of all, only in this paper the singlet is a built-in rather than artificial ingredient. Second, only here the singlet plays a crucial role in EWSB while in others it is just a spectator. EWSB will be one of the centers of our article. Third, only here $N_1$ could enjoy the merits of 
FI$m$P. Last but not the least, we actually will need two scalar singlets for the sake of successful EWSB with acceptable Higgs phenomenologies, and this will make difference in freeze-in.

The paper is organized as follows. In Section II we introduce the $\nu$SISM, studying EWSB, the mass spectrum and freeze-in dynamics. We present the numerical results in section III, and the next section includes  conclusion and discussion. Finally, some supplementary materials are cast in the appendix. 

\section{The scale invariant version of $\nu$SM ($\nu$SISM)}

As stated in the Introduction, the $\nu$SISM predicts the FI$m$P DM $N_1$, which takes advantages in addressing the basic properties of DM such as mass and relic density origins and stability. In this section we will first introduce the model and then study its two main phenomenologies, SI spontaneously breaking and the $N_1$ freeze-in production. The scalar singlets will play crucial roles in both aspects.

We begin with the model setup. Asides from the SM field content, the $\nu$SISM introduces three RHNs $N_i$ to produce realistic neutrino masses and mixings. However, in order to generate Majorana masses for $N_i$, scalar singlet(s) $S_a (a=1,2...n)$ with non-vanishing VEVs are indispensable. Interestingly, they are also badly needed to implement radiative SI spontaneously  breaking. It is found that the minimal case with $n=1$ fails to accommodate the current Higgs data~\cite{HHJ}, so we consider $n=2$ real scalars or a complex singlet $S=(J+i\sigma)/\sqrt{2}$. These minimal degrees of freedom are subject to the $SU(3)_C\times SU(2)_L\times U(1)_Y$ gauge symmetries and the Poincare and classical scale invariance space-time symmetries. Then the most general interacting Lagrangian reads, 
\begin{align}\label{Lag}
-{\cal L}=&\f{\ld_1}{2}|H|^4+\f{\ld_{2}}{2}|S|^4+\ld_3|H|^2|S|^2
+\L\ld_4|H|^2 S^2 +\ld_5 S^3S^*+\f{\ld_6}{2} S^4+c.c.\R.\cr
&+\L\f{\ld_{sn}}{2}SN^2+ y_N\bar \ell H N+c.c.\R.
\end{align}
To reduce parameters, we also impose CP-invariance on the scalar potential, which forces $\ld_{4,5,6}$ to be real. But this symmetry is not physically necessary (we allow complex Yukawa coupling to break it explicitly). 
For later use the expansion of Eq.~(\ref{Lag}) in real degrees of freedom is given in Appendix.~\ref{complex:gen}.

A comment about the novelty of the $\nu$SISM is in order. As a matter of fact, it is not entirely new and similar versions have been investigated in Ref.~\cite{MIS,HHJ}. However, our physical motivations and arguments of the model presented here are quite different to theirs. In addition to that our focused parameter space will be totally different.

\subsection{Scale invariance spontaneously breaking}

\subsubsection{Flat direction and tree level spectrum}

We proceed to investigate the vacuum of the scalar potential given in Eq.~(\ref{Lag}). In the absence of a symmetry, it is supposed that both components of $S$ develop VEVs. So, along with the Higgs doublet VEV, we should work in the three-dimensional field space. Treatment of such a situation becomes more complicated than the single field case, and the Gildener-Weinberg~\cite{GW} instead of the Coleman-Weinberg (CW)~\cite{CW} approach is appropriate. Following this approach, we should work out the flat directions of the scalar potential, namely $\vec \Phi_{{flat}}=\varphi \vec n$ with $\varphi$ and $\vec n$ respectively the modular and directional vector of $\vec \Phi_{flat}$ in the multi-dimensional field space. The existence of such kind of directions is a consequence of SI at tree level, which leads to the modular-independent minimum lines of the potential, i.e., the lines satisfying the equation $dV_{tree}/d\varphi=0$. The flat directions will be lifted by radiative corrections, that introduce sources of SI violation and then fix the value of $\varphi$. We remind that these operations are employed at the renormalization scale $Q$, at which all couplings are inputed. Later, it will be chosen at the SI spontaneously breaking scale.

We have simplified the potential by imposing CP-invariance and this simplification allows us to obtain analytical expressions of the flat direction (and mass spectrum as well). Such a strategy is also adopted in Ref.~\cite{MIS}, but it further imposes a $Z_{4}$ discrete symmetry to eliminate the $\ld_{4}-$ and $\ld_{5}-$terms. The flat direction is the solution of the following three tadpole equations:\begin{align}\label{}
&\f{J^{2}}{h^{2}}=y^2,\\
&\ld_J+3\f{\ld_{hJ}}{y^2}+3x^2\ld_{J\sigma}
=0,\label{T2}\\
&x^3\ld_\sigma+3x\f{\ld_{h\sigma}}{y^2}+3x\,\ld_{J\sigma}
=0,\label{T3}
\end{align}
where we have introduced two VEV ratios $x\equiv \sigma/J$ and as well $y\equiv J/h$. As a convention, we always choose $J$ as the largest scale and thus $y>1$ (we will turn back to it later). The above equations are derived from the potential expressed in terms of real components, see Eq.~(\ref{V:general}) and the definitions of coupling constants Eq.~(\ref{para:re}). Immediately, from these equations one can see that in the $Z_{4}-$symmetric case, where $\ld_4=\ld_5=0$ and thus $\ld_J=\ld_{\sigma},\,\ld_{hJ}=\ld_{h\sigma}$, only solutions $x=\pm1$ are possible. In other words,  $Z_{4}$ forces the flat direction to lie along the special direction $\sigma=J$. But  more generic flat directions can be admitted in the absence of such a symmetry. The resulting Higgs physics is different.

In our later numerical demonstration, VEV ratios $x$ and $y$ will be taken as fixed inputs since they have direct physical meaning. Note that $y$ can be expressed in terms of $x$ and other quartic coupling constants,
\begin{align}\label{}
y^{-2}=-3\f{\ld_{hJ}+x^2\ld_{h\sigma}}{\ld_{h}}.
\end{align}
In all cases of interest we shall find that $y$ is at least moderately larger than 1, and thus there should be a mild hierarchy, i.e., $\ld_{hJ}, \,\ld_{h\sigma}\ll \ld_h$. Otherwise we have to arrange a cancelation between the two terms in the numerator. It is convenient to express the flat direction in terms of $x$, $y$ and $h$ as the following
\begin{align}\label{flat}
\varphi=&R(x,y)h\quad {\rm with}\quad R^2(x,y)=1+y^2+x^2y^2,\\ \label{flat1}
n_h=&\f{1}{R(x,y)},\quad 
n_J=\f{y}{R(x,y)},\quad n_\sigma=\f{xy }{R(x,y)}.
 \end{align}
By definition, we have the relations $v_{h,J,\sigma}=n_{h,J,\sigma} \varphi$. Later we will find that $R(x,y)$ (or $n_h$) has a clear interpretation.

The tree level spectrum along the above flat direction consists of three Higgs states, the Goldstone boson (GSB) of SI spontaneously breaking ${\cal P}$ and two massive states $H_{1,2}$. We should calculate radiative corrections of the tree level potential due to these states. The mixings between these scalars are largely determined at tree level. Concretely, the three states are obtained from the mass squared matrix $M^2$, which in the basis $(h,J,\sigma)$ takes a form of
\begin{align}\label{}
M^2=\f{J^2}{y^2}\times\left(\begin{array}{ccc} \f{\ld_h}{3} & \f{\ld_h}{3y}+{x^2y\,\ld_{h\sigma}}  & -{x\,y\,\ld_{h\sigma}}\\ \quad \quad& \f{\ld_h}{3y^2}+{x^2\ld_{h\sigma}} -x^2y^2\ld_{J\sigma}&  x\,y^2\,\ld_{J\sigma} \\ &  &   -{\ld_{h\sigma} }-y^2\ld_{J\sigma} \end{array}\right)
\end{align}
We have expressed $\ld_{J}$ and $\ld_\sigma$ in favor of others via the tadpole conditions Eq.~(\ref{T2}) and  Eq.~(\ref{T3}), respectively. One can explicitly check that Det$M^2=0$, hence implying a GSB.

The two massive Higgs bosons can be analyzed in a simple way. The Higgs sector must present a quite SM-like Higgs boson $h_{\rm SM}$, which must be dominated by the Higgs doublet, merely containing small fractions of singlets. It means that the Higgs sector can be split into a doublet sector and a singlet sector. In this simplified case, the singlet sector breaks SI through large singlet VEVs, which are then directly mediated to the doublet sector through the mixing term $S^2|H|^2$, generating the negative Higgs mass term for EWSB~\footnote{In this sense, it actually reduces to the Higgs portal-like models with hidden CW mechanism~\cite{SI:hidden}}. Now, the mass squared of $h_{\rm SM}$ can be approximated by
\begin{align}\label{}
m_{h_{\rm SM}}^2=m_{H_1}^2\approx M^2_{11}=\f{\ld_h}{3} v^2.
\end{align}
So as usual $\ld_h\approx 0.75$ is almost fixed by the 125 GeV Higgs boson, up to a small correction from the mixing effect. With this result and taking into account the presence of a GSB, which is dominated by singlets, it is not difficult to deduce the mass squared of the heavier Higgs boson $H_2$:
\begin{align}\label{}
m_{H_2}^2\approx M^2_{22}+M^2_{33}=-\left[(1+x^2)\ld_{J\sigma} +(1-x^2)\ld_{h\sigma}/y^2\right]v_J^2.
\end{align}
In the ensuing discussion we will see that, to guarantee the presence of $h_{\rm SM}$, at least one singlet should develop a larger VEV than that of the doublet, $v$. As mentioned before in our notation we take $v_J=\langle J\rangle\gg v$ or $y=v_{J}/v\gg1$. Then, the mass of $H_2$ is well approximated to be 
\begin{align}\label{H2App}
m_{H_2}\approx  \sqrt{-(1+x^2)\ld_{J\sigma}}v_J.
\end{align}
Or one can express it in terms of $v_{\varphi}$, $ m_{H_2}=x\sqrt{\f{\ld_{\sigma}}{3}+ \f{\ld_{h\sigma}}{x^2y^2}}v_\varphi$, by solving $\ld_{J\sigma}$ from Eq.~(\ref{T3}). In most cases it can be further approximated  to be a very simple form $x\sqrt{\ld_{\sigma}/3}v_\varphi$.

\subsubsection{CW potential and pseudo-GSB mass}

With the massive spectrum at hand, we now discuss SI spontaneously breaking in the one loop CW effective potential $V_{CW}$. In general, it can be written as~\cite{MIS}
\begin{align}\label{}
V_{CW}=A(\vec n)\varphi^4+B(\vec n)\varphi^4\ln \f{\varphi^2}{Q^2}.
\end{align}
The numerical factors $A$ and $B$ are functions of the dimensionless couplings only. In the MS scheme, they are explicitly given by 
\begin{align}\label{}
A=&\f{1}{64\pi^2}\left[\sum_s g_s^4\L-3/2+\ln g_s^2\R+3\sum_V g_V^4\L-5/6+\ln g_V^2\R-4\sum_f g_f^4\L-1+\ln g_f^2\R \right],\cr
B=&\f{1}{64\pi^2}\left(\sum_s g_s^4+3\sum_V g_V^4-4\sum_f g_f^4 \right),
\end{align}
with $s/V/f$ denoting for real scalar/vector boson/Dirac fermion respectively. In writing these expressions, we have collectively assumed the relation $m=g v_\varphi$, which is true in the models with SI. It is worthy of note that in the multi-dimensional field space, if the field that gains mass mainly via coupling to the scalar with a smaller VEV, e.g., $h$ in this context, the resulting effective coupling $g$ will be suppressed by the fraction $n_h\ll1$. In this way, we can avoid the failure of the scale invariant SM, which is caused by the large negative contribution to $B$ from the heavy top quark. We will turn back to this point later.

Within the perturbative region, namely both $A$ and $B$ are much smaller than 1, the VEV of $\varphi$ is pinned down by solving the equation $dV_{CW}/d\varphi=0$. From it we get the relation
\begin{align}\label{vQ}
v_\varphi= Q \exp{\L-A/2B-1/4\R}.
\end{align}
If we choose $Q=v_\varphi$, 
which eliminates the potential large logarithmic terms, we get a relation among the couplings at scale $Q$, i.e., $A+2B=0$. It is nothing but dimensional transmutation~\cite{CW}. Perturbativity of the effective potential is satisfied for $g_s\lesssim 1$. If both $A$ and $B$ receive contributions mainly from a single field dependent mass term, such as $m_{H_2}$ in this paper, we can approximately determine the effective coupling constant to be
\begin{align}\label{vQ}
g_s\approx \sqrt{e}\,Q/v_\varphi,
\end{align}
or $m_s=\sqrt{e}\,Q$. Eq.~(\ref{vQ}) tells that, if we choose  $Q\gg v_\varphi$, then $g_s\gg 1$ and thus it violates purturbativity at $Q$; In the opposite, one may want to choose $Q\ll  v_\varphi$ to get a weak coupling $g_s\ll1$. But it will result in an intolerablely light PGSB with mass suppressed by $Q/v_\varphi$ (or $g_s$). Therefore, $g_s\sim{\cal O}(1)$ is preferred. This fact is useful to find out the favored region of $x$. Recall that 
$m_{H_2}=x\sqrt{\ld_{\sigma}/3}v_\varphi$  (see remarks below Eq.~(\ref{H2App})), thus we have $g_{H_2}\approx x(\ld_{\sigma}/3)^{1/2} $. Then to make $g_s\equiv g_{H_2}\sim 1$ we need $\ld_\sigma\sim 3/x^2$, but it blows up even for $x\lesssim 0.3$. So the vacuum figuration with $x\lesssim 1$ is favored, which means that $H_2$ tends to be a strong mixture of $\sigma$ and $J$. 

Now we turn our attention to the pseudo GSB (PGSB) ${\cal P}$, which is massless at tree level but gets a mass from the effective potential. Before heading towards its mass, we would like to first figure out its doublet fraction $F_{{\cal P}h}$, that is useful in constraining/discovering this PGSB at colliders. After an explicit calculation, one finds that this fraction is nothing but just the doublet projection in the tree level flat direction, see Eq.~(\ref{flat}):
\begin{align}\label{FP}
F_{{\cal P}h}=n_h=1/R(x,y)\ll1,
\end{align}
up to a overall sign. $F_{{\cal P}h}$ is determined solely by two VEV ratios $x$ and $y$. A small $F_{{\cal P}h}$ is necessary not only to hide a fairly light PGSB at LEP but also to keep $h_{\rm SM}$ overwhelmingly dominated by doublet. To sufficiently suppress it, at least one of the singlet VEVs should be significantly larger than $v$, i.e., $y\gg1$ and/or $xy\gg1$. Following the previous convention, we take $y\gg1$ and then we have $F_{{\cal P}h}\approx {1}/{y\sqrt {1+x^2}}$. Note that in this convention $x\leq 1$, thus $xy\gg1$ also means $y\gg1$.

The classical SI is violated by quantum effect which thus generates a mass for the tree level massless GSB. In general, it is given by
\begin{align}\label{}
m_{\cal P}^2=8B v_\varphi^2=\f{1}{8\pi^2}\left(\sum_s g_s^2m_s^2+3\sum_V g_V^2m_V^2-4\sum_f g_f^2m_f^2 \right). 
\end{align}
To ensure that the extreme from $dV_{CW}/d\varphi=0$ is indeed a minimum, $m_{\cal P}^2$ or $B$ must be positive. In the above expression, top quarks have the potential to drive $B<0$ but it is stopped by $H_2$. The stability condition is
\begin{align}\label{}
g_{H_2}^4>12g_t^4 \Rightarrow x^2\L\f{\ld_{\sigma}}{3}+ \f{\ld_{h\sigma}}{x^2y^2} \R>\f{2\sqrt{3}(m_t/v)^2}{y^2\L 1+x^2\R}.
\end{align}
It is actually required that $\ld_\sigma>6.3/y^2x^2(1+x^2)$. No surprise, when $x<1/y\ll1$ (or $x^2y^2\lesssim1$), namely one of the singlet having VEV below the weak scale, one needs a large $\ld_\sigma\sim{\cal O}(10)$ to compensate the relatively larger suppression by $n_\sigma$ in $g_{H_2}$ (compared to that by $n_h$ in $g_t$). This is not an appealing situation if we want to keep the model perturbative up to a very high scale. Moreover, the resulting spectrum, in particular ${\cal P}$, is fairly light and thus may have already been excluded by the present experiments like LEP and LHC~\footnote{Practically such a VEV pattern will make for freeze-in RHN in the single RHN limit, and at some corner of the parameter space the light Higgs bosons are also allowed. But we only consider the bulk space, with clear and safe Higgs phenomenologies.}. Therefore,  
we will focus only the case $x^2y^2\gg1$, consistent with the analysis below Eq.~(\ref{vQ}) that concludes $x$ should be near 1. In the $H_2-$dominance limit, after using Eq.~(\ref{vQ}) the PGSB mass is expressed to be the following forms
\begin{align}\label{}
m_{\cal P}=\f{g_s^2}{2\sqrt{2}\pi}v_\varphi=\f{\ld_\sigma/3+\ld_{h\sigma}/x^2y^2}{2\sqrt{2}\pi}x^2\sqrt{1+x^2} \,yv \approx 
0.038\,{\ld_\sigma}\,x^2\sqrt{1+x^2}\, y  v.
\end{align}
As an estimation, we write $m_{\cal P}=75.7\times(y/5.0)(\ld_\sigma/2.0)\,\rm GeV$ with $x=0.8$ fixed. Increasing $y$ helps to not only  suppress the doublet fraction of ${\cal P}$ but also enhance the mass of ${\cal P}$, thus making it safe under the LEP exclusion. The price is pushing $H_2$ into the TeV region, thus hard to detect at near future colliders. But ${\cal P}$ is still promising. We leave more quantitive analysis in the coming section and in the ensuing subsection we enter into the discussions about dark matter, the FI$m$P $N_1$.

\subsection{Freeze-in Sterile Neutrino }

Before heading towards the freeze-in production of sterile neutrino dark matter $N_1$, we briefly discuss its conventional production mechanisms. The tiny active neutrino mass leads to a naive upper bound on the Yukawa coupling constant given in Eq.~(\ref{Lag}) (the bound may be spoiled somehow in the presence of flavor structure),
\begin{align}\label{yN}
y_N\lesssim 10^{-10}\L\f{m_\nu}{0.1\,\rm eV}\R^{1/2}\L\f{M_{N_1}}{10\,\rm keV}\R^{1/2}.
\end{align}
That feeble coupling means that $N_{1}$ never enters the thermal plasma, given no other interactions. It can be non-thermally produced via non-resonant sterile-active neutrino oscillation, known as the Dodelson-Widrow (DW) mechanism~\cite{Dodelson}. But it has been ruled out (we will give reasons later). The resonant production~\cite{SF} and thermal production mechanisms~\cite{Bezrukov:2009th} are still allowed. However, both suffer some theoretical defects since they require big modifications. The former requires an anomalously large lepton asymmetry, and the latter requires a large entropy release. Model extensions are then unavoidable~\footnote{If the two heavier RHNs have quasi degenerate masses, typically with degeneracy $\lesssim 10^{-4}$, they are even capable of generating the observed baryon asymmetry of the Universe~\cite{Canetti:2012kh}. But generically it implies a large fine-tuning.}.

In the $\nu$SISM new interactions of $N_1$ with the singlets appear naturally. These new interactions originally are supposed to generate very light mass for $N_1$ (lightness is necessary for stability), but at the same time they surprisingly can account for correct relic density of $N_1$ via the freeze-in mechanism.

\subsubsection{A toy analysis }

As a toy analysis, only one singlet is considered for the time being. For the keV scale $N_1$, again for stability (and for cosmological consideration~\cite{Weinberg:2013aya}), the strength of the coupling is extremely small $\ld_n\sim10^{-8}$. Hence this new vertex can not thermalize $N_1$ either. But the magnitude of $\ld_n$ is just at the correct order to admit freeze-in production of $N_1$. In this scenario, $N_1$ is produced via the slow decay process $S\ra N_1N_1$, which has negligible inverse decay rate. Note that generically the annihilation processes like $f\bar f\ra \bar N_1N_1$ contribute to the freeze-in process sub-dominantly~\cite{Chu:2011be}, due to the reason, among others, that they are suppressed by extra small couplings. The freeze-in process lasts until the mother particle decouples from the plasma and quickly decays away. The peak of production is around the mass scale of the mother particle $S$. In other words, it is UV-insensitive (for UV-sensitive freeze-in, please see Ref.~\cite{UVF}). For freeze-in proceeding via two-body decay  $P\ra \bar XX$, the final yield of $X$ is formulated to be~\cite{freezein,Kang:2010ha,Chu:2011be,Yaguna}
\begin{align}\label{freezein}
Y_X(\infty)\approx{45\,g_P\over1.66 \pi^4g_{*}^S \sqrt{g_{*}^\rho}}\f{\Gamma(7/2)\Gamma(5/2)}{16}\f{M_{\rm Pl}}{m_{P}^2}\Gamma(P\ra \bar XX),
\end{align}
with  $g_{*}^{S,\rho}$ respectively the effective numbers of degrees of freedom for the entropy and energy densities at $T\simeq m_P$, the mass of the mother particle. $g_P$ is the internal degrees of freedom of $P$. For multi mother particles contributing to freeze-in $X$, there is a summation over $P$. Specified to this schematic example for freeze-in $N_1$, the relic density depends on the unknown parameters as proportional to $\ld_n^3 \langle S\rangle$. Then, for a TeV scale $ \langle S\rangle$, it is found that a keV scale FI$m$P allows for correct relic density. This kind of feeble interaction admitting correct relic density of extremely light DM 
is somewhat reminiscent of weak interaction admitting correct relic density of weak scale DM. Here TeV scale, a sign of new physics, is also involved, but it is in the form of VEV rather than DM mass scale itself. 

Analysis in the realistic model becomes more complicated from two aspects. First, there are two singlets  $J$ and $\sigma$ coupling to $N_1$, both having VEVs. Moreover, two physical Higgs bosons ${\cal P}$ and $H_{2}$ (actually three but the contribution from the SM-like Higgs boson is suppressed by small mixing) contribute to the freeze-in process. Second, it is well known that a single family of RHN fails to accommodate neutrino phenomenologies, and thus a nontrivial flavor structure should be taken into account. This, along with the multi singlets, is going to make a big difference. We find that there are two distinguishable scenarios that can successfully freeze-in $N_1$, and one of them is just reduced to the single RHN case.

\subsubsection{Consider multi-singlets $\&$ flavor structure}

As a warm up, we consider the case with only one RHN $N_1$ but two singlets. The Lagrangian relevant to freeze-in is derived as the following,
\begin{align}
\f{\eta_{S_\alpha }}{2}S_\alpha N_1  N_1\ra \f{M_{N_1}}{2}N_1^2+\f{\eta_{H_a}}{2} {H_a} N_1N_1,\quad \eta_{H_a}\equiv \sum_\alpha\eta_{S_\alpha}F_{S_\alpha H_a},
\end{align}
with $S_\alpha=(J,\,\sigma)$ and $M_{N_1}=\eta_{S_\alpha}v_{S_\alpha}=(x\,\eta_J+\eta_\sigma)v_\sigma$~\footnote{The SM Higgs doublet also contributes to mass of $N_1$ via the dimension-five operator $|H|^2N_1^2/\Ld$, which is obtained after integrating out the singlet $S$ with VEV $v_s$ and mass $m_S$. Roughly, $\Ld\sim \ld_{sh} \ld_{sn}v_s/{m_S^2}\sim \f{\ld_{sn}}{v_s} \L m_h/m_S\R^2$, where we have used the induced weak sale from $\ld_{sh}|H|^2S^2$. Therefore, compared to the contribution from $S$ to $N_1$, $M_{N_1}\sim \ld_{sn}v_s$, this contribution is suppressed by $\L\f{v}{v_s} \f{m_h}{m_S}\R^2\ll 1$.}. We have written $S_\alpha=F_{S_\alpha H_a}H_a$ with $H_a=({\cal P},\,H_1,\,H_2)$. In the interested region with $x\lesssim 1$
and $y\gg1$, the $H_1$ ($=h_{\rm SM}$) component can be neglected. While other fractions are approximately parameterized by one mixing angle $\theta$ between the singlets: $F_{J{\cal P}}=F_{\sigma H_2}=\cos\theta$ and $-F_{JH_2}=F_{\sigma{\cal P}}=\sin\theta$ with $\tan\theta=x$ and $\theta\in[0,\pi/4]$. It is illustrative to write $\eta_{H_a}=f_{H_a}{M_{N_1}}/{v_J}$ with 
\begin{align}
f_{\cal P}= \left[1+\f{v_J}{M_{N_1}}(1-x^2) \eta_J\right]\cos\theta,\quad f_{H_2}=\f{1}{x} \left(1-2x\f{v_\sigma}{M_{N_1}}\,\eta_J\right)\cos\theta.
\end{align}
Barring cancellation between the two contributions to $M_{N_1}$, one gets the naive estimations $\eta_{\sigma,J}\sim M_{N_1}/v_{\sigma,J}$. In particular, if only one singlet couples to $N_1$, there will be no cancelation and then for $x\simeq1$ it is expected $f_{H_\alpha}\sim 1$, the reference value for $f_{H_\alpha}$ hence. Now substituting the decay width of ${H_a}\ra N_1N_1$ into Eq.~(\ref{freezein}) one can estimate the DM relic density $\Omega_{\rm DM} h^2=2.82\times10^2\L\f{m_{\rm DM}}{\rm keV}\R Y_{\rm DM}(\infty)$ as
\begin{align}\label{relic}
\Omega_{\rm DM} h^2=0.11\times\sum_{H_a={\cal P},H_2}\L\f{f_{H_a}^2}{1.0}\R
 \L\f{m_{\rm DM}}{10\,\rm keV}\R^3\L\f{\rm TeV}{v_J} \R^2\L\f{100\,\rm GeV}{m_{H_a}}\R\L \f{10^3}{g_{*}^S \sqrt{g_{*}^\rho}}\R,
\end{align}
with $m_{\rm DM}=M_{N_1}$. 

From the above equation we find that there is a mild tension between the Higgs sector and the dark sector. And the tension gets more serious as $N_1$ becomes lighter, e.g., as light as the potential warm dark matter with mass around 1 keV. For that light DM, Eq.~(\ref{relic}) shows that the dark sector wants the singlets' VEV to lie significantly below the TeV scale, which however is disfavored by the Higgs phenomenology in terms of the previous discussions. Therefore, we may have to endure a substantial cancelation between the two singlets coupling to $N_1$, so as to make at least one $f_{{H_a}}\sim10$. Immediately, we know that the case with only one singlet coupling to RHN fails in freezing-in a quite light ${N_1}$, because it always gives $f_{{H_a}}\sim1$.

Now let us detail how incorporating the flavor structure for RHNs makes a big difference. Actually, it opens up a novel scenario for freeze-in. To see it, we consider a simplified case, i.e., there is a large mass hierarchy between the DM candidate and the extra RHNs, for instance, the other two RHNs lying at the GeV scale inspired by baryogenesis~\cite{Canetti:2012kh}. The genetic Yukawa couplings are $\eta_{S_\alpha,ij}S_\alpha N_iN_j/2+c.c.$ with $\eta_{S_\alpha,ij}=\eta_{S_\alpha,ji}$ and $i/j=1,2,3$. After writing $S_\alpha\ra v_{S_\alpha}+S_\alpha$, we as usual can work in the mass eigenstates of RHNs through an unitary rotation $N_i\ra U_{ij}N_j$ and eventually arrive at the Lagrangian
\begin{align}
-{\cal L}_{N}=\f{M_i}{2}N_i^2+\f{\eta_{H_a,ij}}{2} {H_a} N_iN_j+c.c.,
\end{align}
with $M_{N_1}\ll M_{N_2}\leq M_{N_3}$ and $\eta_{H_a,ij}=F_{S_\alpha H_a}(U^\dagger)_{ii'}\eta_{S_\alpha,i'j'}U_{j'j}$. Owing to multi singlets with VEVs, the mass matrix and the Yukawa coupling matrix of RHNs can not be diagonalized simultaneously. Consequently the interaction of $N_1$ is more involved than that of the previous case, which is only the limit of negligible mixing effect in the case considered here. To understand this limit better and for later convenience, we turn back to the original mass matrix $M_N$ (for illustration only two RHNs are considered): 
\begin{align}
M_N=\left(\begin{array}{cc} \eta_{J,11}v_J+\eta_{\sigma,11}v_\sigma & \eta_{J,12}v_J+\eta_{\sigma,12}v_\sigma  \\\eta_{J,12}v_J+\eta_{\sigma,12}v_\sigma &   \eta_{J,22}v_J+\eta_{\sigma,22}v_\sigma
\end{array}\right)\sim \left(\begin{array}{cc}{\cal O}(10^{-6}) &  {\cal O}(10^{b}) \\ {\cal O}(10^{b}) & {\cal O}(10^{2a})\end{array}\right)\rm\,GeV, 
\end{align}
with $-3\lesssim a\lesssim 0$. To make the lighter eigenvalue naturally $\sim {\cal O}(10^{-6})$ GeV, $b$ should be smaller than $a-3$. The single RHN case discussed before corresponds to $b\ll a-3$, a condition to decouple the heavy from the light.

On the contrary, when $b\lesssim a-3$ the light-heavy RHN mixing effect is not negligible and can play an important role. Still consider the hierarchical scenario, where the mixing angle is estimated to be $\sin\theta_N\simeq (M_{N})_{12}/(M_N)_{22}\sim 10^{b-2a}\ll1$. Two off diagonal Yukawa coupling constants $\eta_{H_a,12}$, by naive estimation, are approximated to be
\begin{align}
\eta_{H_a,12}\approx & F_{S_\alpha H_a}\left(\eta_{S_\alpha,11}\sin\theta_N+\eta_{S_\alpha,12}-\sin^2\theta_N\eta_{S_\alpha,21}-\sin\theta_N\eta_{S_\alpha,22}\right)\cr
\simeq & F_{S_\alpha H_a}\L \eta_{S_\alpha,12}- \sin\theta_N\eta_{S_\alpha,22}\R.
\end{align}
After some exercise, it is not difficult to find that at leading order $\eta_{{\cal P},12}=0$. But the one involving $H_2$ is not zero, and explicitly it is given by 
\begin{align}
\eta_{H_2,12}\approx \cos\theta \left(\eta_{\sigma,12}-x\,\eta_{J,12}\R - \cos\theta  \left(\eta_{\sigma,22}-x\,\eta_{J,22}\R
\f{x\,\eta_{\sigma,12}+\eta_{J,12}}{x\,\eta_{\sigma,22}+\eta_{J,22}}.
\end{align}
As one can see, $\eta_{H_2,12}\sim \eta_{S_\alpha,12}\sim 10^{b}/v_{S_\alpha}$. It is $b-6$ orders of magnitude larger than $\eta_{H_a,11}$, which is expected to be $\sim10^{-9}$ for a keV RHN. Recall that $b<a-3$, thus as long as $a$ is not less than $-2$, $\eta_{H_2,12}$ can be easily around $10^{-8}$. In the light of the estimation in Eq.~(\ref{relic}), this light-heavy mixing effect is able to freeze-in a quite light ($\sim$keV) $N_1$ via decay $H_2\ra N_1N_2$. We stress again that the presence of multi singlet scalars is a key to preserve a significant mixing effect $\eta_{H_a,12}\neq0$. In summary, in our paper correct relic density of $N_1$, no matter light or heavy, can be readily achieved by means of freeze-in.

\section{The numerical analysis of the $\nu$SISM}

In this section we investigate the numerical results of the Higgs and dark sectors phenomenologies, respectively. For the former, we are not interested in making a parameter space scan. Instead, we try to make the features of the Higgs boson spectrum visual. For the latter, we even do not involve the original parameters, since from the previous analysis it is known that, the correct relic density of $N_1$ can always be achieved over a fairly wide region of $M_{N_1}$. So the dark matter phenomenologies actually involve only two parameters, $M_{N_1}$ and $\sin\theta_1$, the mixing angle between $N_1$ and active neutrinos. Noticeably, the recently reported $X-$ray line at 3.55 keV can be readily explained by the decaying $N_1$ (7.1 keV DM with freeze-in production to explain this line was also considered in Ref.~\cite{Xfreezein}). While it is problematic for $N_1$ produced by other mechanisms.

\subsection{On the Higgs bosons}

The analysis of the Higgs sector includes a theoretical constraint, i.e., the potential should be bounded from below (BFB) in order to make the minimum stable. We check this at tree level. It is more convenient to do it in the three-dimensional field space spanned by $(h,J,\sigma)$, where the potential can be written as $V=X \Ld X^T$ with $X=(h^2,J^2,\sigma^2)$ the bilinear vector and the matrix of coupling constants
\begin{align}\label{}
 \Ld=\f{1}{8}\left(\begin{array}{ccc}  \ld_h/3 & \ld_{hJ} & \ld_{h\sigma} \\  \ld_{hJ}& \ld_J/3&  \ld_{J\sigma}  \\ \ld_{h\sigma} & \ld_{J\sigma} & \ld_\sigma/3\end{array}\right).
\end{align}
BFB requires that all of its sub-matrices $\Ld_{n\times n}$ with $n=1,2,3$ should satisfy the conditions ${\rm Tr}\Ld_{n\times n}>0$ and ${\rm Det}\Ld_{n\times n}>0$. Equivalently, the following conditions for the quartic couplings should be fulfilled, 
\begin{align}\label{BEF}
&\ld_h>0, \quad \ld_J>0,\quad \ld_\sigma>0,\cr
&|\ld_{hJ}|<\f{1}{3}\sqrt {\ld_h\ld_J},\quad |\ld_{h\sigma}|<\f{1}{3}\sqrt {\ld_h\ld_\sigma},\quad |\ld_{J\sigma}|<\f{1}{3}\sqrt {\ld_J\ld_\sigma},\cr
&|\ld_h\ld_{J\sigma}/3-\ld_{hJ}\ld_{h\sigma}|<\sqrt{\ld_h\ld_J/9-\ld_{hJ}^2}\sqrt{\ld_h\ld_\sigma/9-\ld_{h\sigma}^2}.
\end{align}
Basically, they are requiring sufficiently small off-diagonal elements in $\Ld$. Note that they practically lead to mass mixings in the Higgs mass matrix,  thus the above conditions are consistent with the requirement that there should be a quite SM-like Higgs boson. 

\begin{figure}[htb]
\begin{center}
\includegraphics[width=0.42\textwidth,natwidth=610,natheight=642]{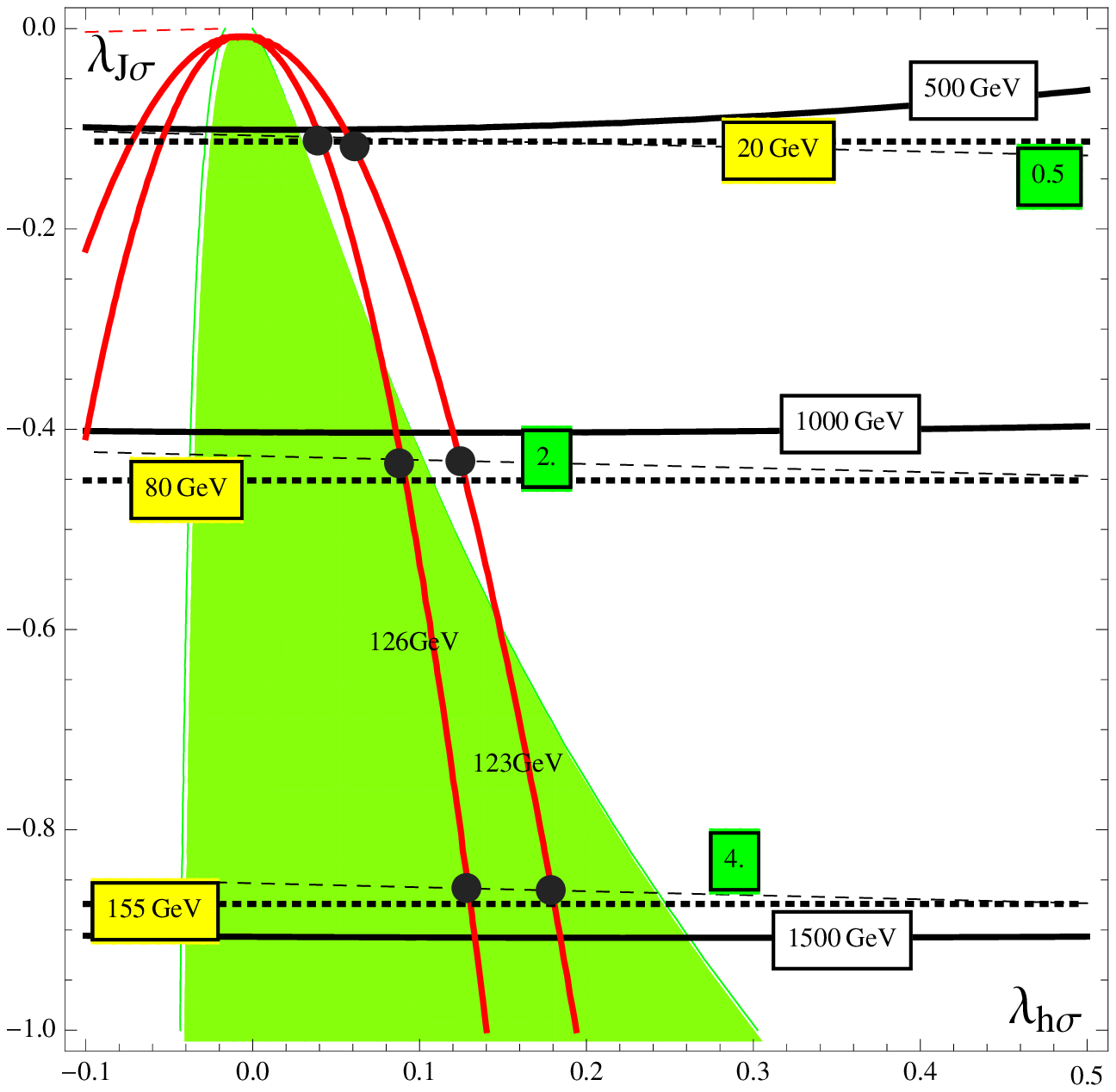}~~~~~~~~~
~~~
\includegraphics[width=0.42\textwidth,natwidth=610,natheight=642]{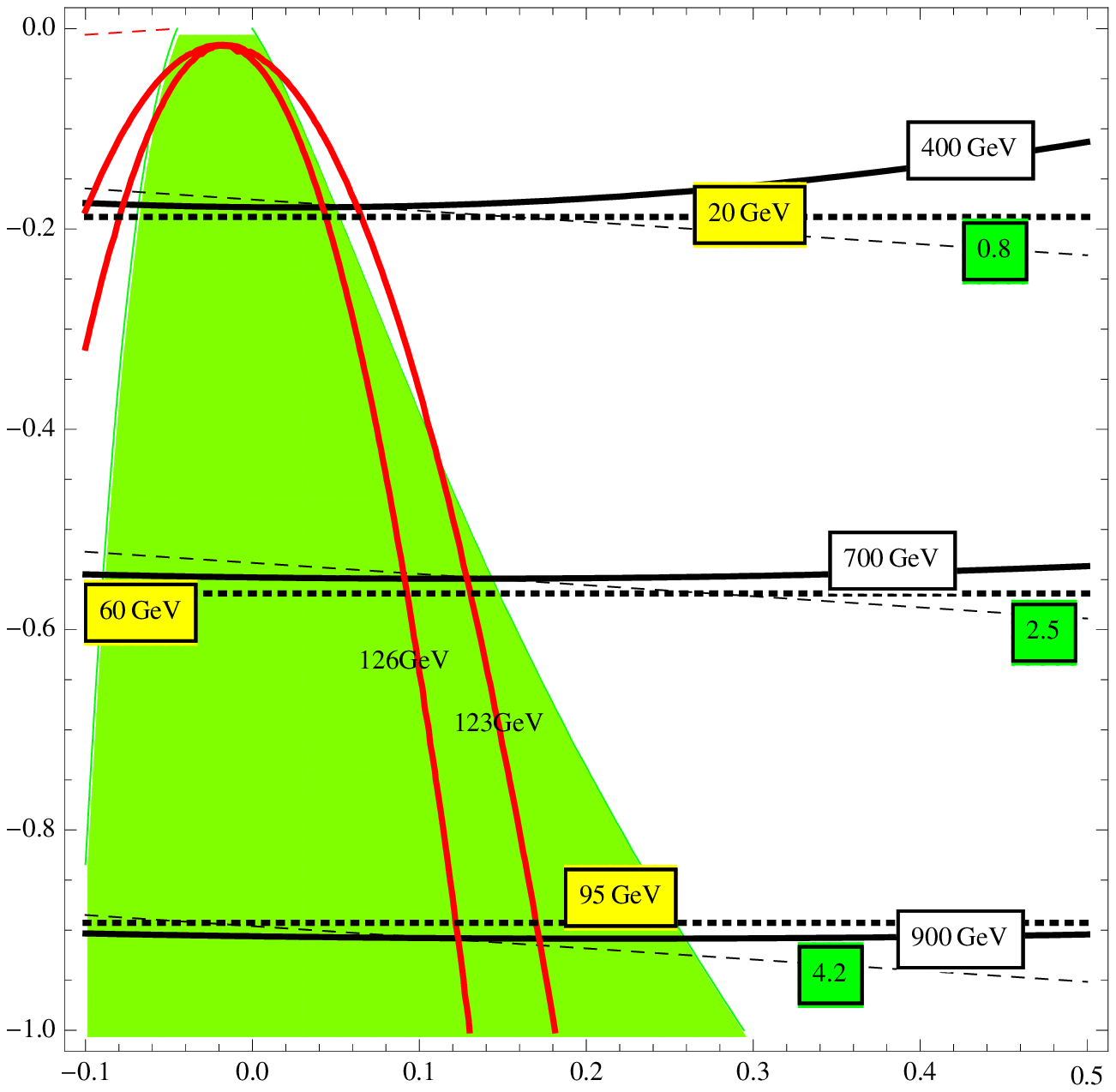}
\caption{Left: Contour plots of masses of the SM-like Higgs boson (red thick lines, two values 123 GeV and 126 GeV taken), heaviest Higgs boson $m_{H_2}$  (black thick lines) and PGSB  (dotted thick). The dashed lines are for different choices of the values of $\ld_\sigma$: 0.5, 2, 4. The green-shadowed region has been excluded by the BEFB condition. We take $y=5$ and $x=0.8$, $\ld_h=0.78$. Right: The same but with $y=3$ for comparison.
\label{xsec}}
\end{center}
\end{figure}
Aside from the renomalization scale $Q$, altogether the Higgs sector contains six real parameters, $\ld_{h,\sigma,J}$ and $\ld_{h\sigma}$,  $\ld_{hJ}$,  $\ld_{J\sigma}$. Among them, $\ld_h$ is almost fixed by the SM-like Higgs boson mass, and three can be expressed in terms of two VEV ratios $x, \,y$ and $\ld_{h\sigma},\,\ld_{J\sigma}$. Since $x$ and $y$ have clear physical implications, they are taken to be inputs and fixed in our numerical samples. Finally only three free parameters $\ld_\sigma$, $\ld_{h\sigma}$ and $\ld_{J\sigma}$ are left. We study the Higgs spectrum varying with them and display the results in Fig.~\ref{xsec}, on the $\ld_{h\sigma}-\ld_{J\sigma}$ plane, with $\ld_\sigma$ chosen schematically. For each chosen $\ld_\sigma$ (dashed line), we show the corresponding mass spectrum through two lines, the thick line and dotted line for $m_{H_2}$ and $m_{\cal P}$, respectively. For comparison, in  Fig.~\ref{xsec} we show two choices of $y$: $y=5$ (left panel) and $y=3$ (right panel). Some observations are available from the figure:
\begin{itemize}
\item In the singlet-doublet decoupling limit, $m_{h_{\rm SM}}$ in expectation is merely sensitive to the diagonal quartic coupling constant $\ld_{h}$. Similarly, masses of $H_2$ and ${\cal P}$ are sensitive to the diagonal quartic coupling constant in the singlet sector $\ld_{\sigma}$ but insensitive to the singlet-doublet mixing coupling constant $\ld_{h\sigma}$. 
\item For all of the BEFB conditions listed in Eq.~(\ref{BEF}), practically the last one suffices. It excludes the regime outside the green-shadowed area. The intersection (black circle) between the Higgs mass- and $\ld_\sigma$-curve is called a solution, determining a set of parameters $(\ld_\sigma,\ld_{h\sigma},\ld_{J\sigma},...)$. BEFB is able to rule out the smaller values of $\ld_\sigma$ for a larger $y(=5)$, but for the smaller $y(=3)$ this power tends to be lost. For illustration, in the $y=5$ case we show three typical solutions with $\ld_{\sigma}=0.5,\,2.0$ and 4.0, respectively.  One can see that given $m_{h_{\rm SM}}=123-126$ GeV, the first one has been excluded by BEFB,  the second one is near exclusion while the last one is safe. 
 
\item The region giving $m_{\cal P}<114.4$ GeV is subject to the LEP constraint and $(m_{\cal P},\, F_{{\cal P }h})$ are restricted. In turn, $\ld_{\sigma}$ and $y$ are bounded. We do not intend to scan the whole parameter space since the studies of the SM Higgs doublet mixing with a singlet scalar have been done in many works, say, the most relevant one~\cite{HHJ}. Generically, as long as $\ld_\sigma$ is around 1 and $y$ is relatively large, ${\cal P}$ can easily avoid the LEP bound; if and only if $y$ and $\ld_{\sigma}$ are of normal size, it is hopeful to hunt at least one of the two extra Higgs bosons. If in the future we do really hunt two new Higgs bosons with hierarchical masses, the $\nu$SISM will be a good candidate to account for them. 
\end{itemize}

\subsection{On the FI$m$P with a benchmark at 3.5 keV $X-$ray line}

\begin{figure}[htb]
\begin{center}
\includegraphics[width=0.6\textwidth,natwidth=610,natheight=642]{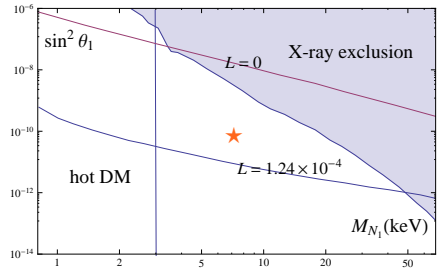}
\caption{In this plot, the shadowed region is excluded by the $X-$ray observations, and the vertical line at $M_{N_1}=3$ keV indicates the lower bound on $M_{N_1}$ from the free streaming limit. On the line with lepton asymmetry $L=0$, correct relic density of $N_1$ is achieved via the DW mechanism; While on the line with  $L=1.24\times 10^{-4}$, it is via resonant production (we use data from Ref.~\cite{Canetti:2012kh}).  The red star labelling the point that fits the 3.55 keV $X-$ray anomaly. It lies in the bulk parameter space of the freeze-in scenario. 
\label{bounds}}
\end{center}
\end{figure}

Before heading towards the freeze-in production, we explain why $N_1$ with DW mechanism has been ruled out already. $N_1$ can decay into a neutrino plus photon, with decay width~\cite{osci}
\begin{align}\label{}
\Gamma_{N_1\ra \nu\gamma}\simeq \f{9G_F^2\alpha M_{N_1}^5}{256\pi^4}\times\sum_{\alpha} \sin^2\theta_{\alpha1},
\end{align}
where $\theta_{\alpha1}$ is the mixing angle between $N_1$ and the active neutrino flavor $\nu_\alpha$. The nonobservation of $X-$ray line stringently constrains on $M_{N_1}$ and $\sum_{\alpha} \sin^2\theta_{\alpha1}\equiv \sin^2\theta_1$~\cite{Xrayb}. The width is proportional to $M_{N_1}^5$, so $\theta_1$ is restricted to be very small for a heavier $N_1$. Consequently, the final yield of $N_1$ via the DW production is insufficient~\cite{osci}:
\begin{align}\label{}
\Omega_{\rm DW}h^2\approx 0.016\times \L\f{\sin^2\theta_1}{10^{-10}}\R\L\f{M_{N_1}}{5\,\rm keV}\R^{1.8}.
\end{align}
Concretely, the region $M_{N_1}\gtrsim3-4$ keV has been excluded. On the other hand, the Lyman-$\alpha$ (Ly$\alpha$) forest gives a compensatory constraint on the lighter RHN which has a longer length of free streaming $\ld_{fs}$. The latest analysis yields $M_{N_1}\gtrsim8$ keV~\cite{Lalpha}. Therefore, the entire region of $M_{N_1}$ has been excluded, see Fig.~\ref{bounds}.

The DM production via freeze-in changes the situation dramatically. Firstly, the $X-$ray bound can be avoided because the freeze-in production mechanism has nothing to do with the sterile-active neutrino mixing. In principle, it can be arbitrarily small given three families of RHN, because even in the limit of decoupling $N_1$ from the active neutrinos the other two heavy RHNs can still account for neutrino masses and mixing~\footnote{Such a limit amounts to the singlet fermonic FI$m$P via a singlet scalar portal~\cite{Yaguna}. Obviously, in that limit one does not need to worry about the $X-$ray bound.}. Secondly, the Ly-$\alpha$ bound relaxes significantly because of two reasons. One is that $f(p,t)$, the initial spectrum of $N_1$, becomes slightly colder in the freeze-in scenario, where the average momentum of $f(p,t)$ is $\langle p_T\rangle\approx 2.45$ T, while in the DW scenario $\langle p_T\rangle\approx 2.8$ T~\cite{inflaton}. The other one is due to a significant entropy dilution. 
Here $N_1$ is produced at the very early Universe $t_{in}$, corresponding to the temperature $T_{in}\simeq m_{H_a}\sim 0.1-1$ TeV. From $t_{in}$ to the current time $t_0$ there is a large entropy release ${\cal S}\equiv g_{*}(t_{in})/g_{*}(t_{0})$, which substantially cools $N_1$ down. Eventually, the average momentum becomes
\begin{align}\label{}
{\cal S}^{-1/3}\langle p_T\rangle\approx {\cal S}^{-1/3} 2.45\, T.
\end{align}
The Ly-$\alpha$ bound accordingly weakens and merely gives $M_{N_1}>$1 keV, a fairly loose bound. But to avoid a hot $N_1$ yields a stronger lower bound. In terms of Eq.~(\ref{FS}) the free streaming of $N_1$ is estimated to be~\footnote{Different to the freeze-in scenario through a frozen-in scalar~\cite{Merle:2013wta,Adulpravitchai:2014xna}, here the length of free streaming is independent on mass of the decaying scalar boson.}
\begin{align}\label{}
\ld_{fs}={\cal S}^{-1/3}\f{\sqrt{a_{eq}}}{t_{eq}}\sqrt{t_{nr}}\L5+\ln\L\f{t_{eq}}{{t_{nr}}}\R\R\simeq 0.038\times\L\f{100}{g_*}\R^{1/4}\L\f{10\rm keV}{M_{N_1}}\R\L\f{33}{\cal S}\R^{1/3}{\rm\,Mpc}.
\end{align}
where $\sqrt{t_{nr}}$ is the time at which RHN becomes nonrelativistic. To ensure that $N_1$ does not turn out to be hot, one requires $\ld_{fs}\lesssim 0.1$ Mpc and in turn $M_{N_1}\gtrsim 3$ keV.

Interestingly, a $X-$ray line at energy 3.55 keV is recently reported with 3$\sigma$ significance evidence~\cite{Boyarsky:2014jta}, through the observation of galaxy clusters and the Andromeda galaxy. Despite of the controversy~\cite{banana}, it is tempting to interpret it as a smoking gun of decaying sterile neutrino with mass about 7.1 keV and mixing angle $\sin^22\theta_1\approx 7\times 10^{-11}~$\cite{Boyarsky:2014jta}. Production mechanisms of correct relic density for such $N_1$ are of concern. The non-resonant production fails already. The resonant production mechanism may work~\cite{Xray:res}, but the latest work Ref.~\cite{Merle:2014xpa} showed that it also fails after taking into account the Ly-$\alpha$ bound. While the freeze-in mechanism that gives a colder RHN,  either via a frozen-in~\cite{Adulpravitchai:2014xna,Merle:2014xpa} or a thermal scalar boson decay, successfully accommodates that $N_1$ with correct relic density in the bulk parameter space, see Fig.~\ref{bounds}.

 \section{Conclusion and discussion}

We proposed the FI$m$P framework for dark matter, a combination of FIMP with SI. It is consistent with the null results from all kinds of DM detections. Besides, it shows advantages in addressing basic questions about DM, stability, mass origin and relic density generation, in an inherently way. In the golden example, the $\nu$SISM, the FI$m$P candidate the lightest RHN $N_1$ furthermore is predicted instead of introduced artificially. We would like to stress that, another attraction of the $\nu$SISM is its economic and self-consistence. Owing to scale invariance, it is necessary to incorporate extra singlets that develop VEVs around TeV scale to generate Majorana masses for the RHNs; At the same time, they are capable of producing $N_1$ via freeze-in, addressing its relic density problem. Not only that, these singlets are badly needed for SI spontaneously breaking itself.

Two open questions deserve further exploration. Firstly, in this paper we actually work in the three RHNs scenario, so the $X-$ray bound is simply gone in the decoupling limit of  $N_1$. But it is tempting to work in a more predictive framework where $N_1$ plays a more active role in neutrino physics, e.g., only two RHNs are introduced and then $N_1$'s coupling to active neutrinos can not decouple. In that case, the resulting $X-$ray line will be closely correlated with neutrino phenomenologies. Secondly, it is of special interest to explore sterile neutrino DM in the scale invariant $B-L$ models~\cite{SIBL1,SIBL} where the RHNs have more natural physical origin, i.e., they are required by anomaly cancelation. But in such kinds of models RHNs carry $B-L$ charge and thus they are thermalized, except in the limit of decoupling new gauge dynamics, e.g., the new gauge coupling is vanishingly small or the massive gauge boson is very heavy such that it decouples before reheating.

   \section{Acknowledgements}
We would like to thank Xiaoyong Chu and P. Ko for helpful discussions.

\appendix
\section{The most general scalar potential: from complex to real}\label{complex:gen}

For the Higgs potential that consists of the SM Higgs doublet plus a complex singlet $S=(J+i\sigma)/\sqrt{2}$ and respects SI, its most general form reads
 \begin{align}\label{}
V_{general}=&\f{\ld_1}{2}|H|^4+\f{\ld_{2}}{2}|S|^4+\ld_3|H|^2|S|^2
+\L\ld_4|H|^2 S^2 +\ld_5 S^3S^*+\f{\ld_6}{2} S^4+c.c.\R.
\end{align}
It contains three real and three complex quartic coupling constants. One can also rewrite it in terms of the three real (physical) degrees of freedom, $(h,J,\sigma)$:
\begin{align}\label{V:general}
V_{general}=&\f{\ld_{1}}{8}h^4+\f{1}{4}\L \f{\ld_2}{2}-2{\rm Re} \ld_5+{\rm Re}\ld_6\R J^4+\f{1}{4}\L \f{\ld_2}{2}+2{\rm Re} \ld_5+{\rm Re}\ld_6\R \sigma^4\cr
&+\f{1}{4}    \L\ld_3-2{\rm Re}\ld_4\R h^2J^2 +\f{1}{4}    \L\ld_3+2{\rm Re}\ld_4\R h^2\sigma^2 +
\f{1}{4}    \L\ld_2-6{\rm Re}\ld_6\R J^2\sigma^2 \cr
&   -{\rm Im}(\ld_5-\ld_6) J^3\sigma-{\rm Im}(\ld_5+\ld_6) J\sigma^3-{\rm Im}\ld_4\,h^2 J\sigma.
\end{align}
Imposing CP-invariance on the Higgs sector forces the CP-odd part $\sigma$ to appear in pair and then terms in the last line disappear. In this sense, CP-invariance is equivalent to an $Z_2$ acting on $\sigma$. Of course, SI accidentally renders $J$ and $h$ charged under other accidental $Z_{2}$. Additionally, if we define 
\begin{align}\label{para:re}
&\ld_h\equiv 3\ld_1,\quad \ld_J\equiv 3{\ld_2}-12 \ld_5+6{\rm Re}\ld_6,\quad \ld_\sigma\equiv 3{\ld_2}+12 \ld_5+6{\rm Re}\ld_6,\cr
& \ld_{hJ}\equiv \ld_3-2{\rm Re}\ld_4,\quad \ld_{h\sigma}\equiv \ld_3+2{\rm Re}\ld_4,\quad \ld_{J\sigma}\equiv \ld_2-6{\rm Re}\ld_6,\cr
&\ld_7\equiv -6{\rm Im}(\ld_5-\ld_6) ,\quad \ld_8\equiv -6{\rm Im}(\ld_5+\ld_6),\quad \ld_9\equiv  -2{\rm Im}\ld_4,
\end{align}
the potential can be written as the form of $V=\ld_{mml}h^nJ^m\sigma^l/n!m!l!$ with $n+m+l=4$.

\section{Free streaming scale}\label{reestream}

A particle after decoupling from the thermal bath travels freely within the gravity potential. From the production time $t_{in}$ of the particle to the present time $t_0$, which is far later than the matter radiation equality time scale $t_{eq}=10^{11}s$, the mean free scale, i.e., the free streaming scale can be calculated via~\cite{Merle:2013wta}
\begin{align}\label{}
\ld_{fs}=\int_{t_{in}}^{t_0}\f{v(t)}{R(t)}dt=\int_{t_{in}}^{t_{nr}}\f{1}{R(t)}dt+\int_{t_{nr}}^{t_{eq}}\f{v(t)}{R(t)}dt+\int_{t_{nr}}^{t_0}\f{v(t)}{R(t)}dt,
\end{align}
We have divided the integral into three regions. In the first region, particle is relativistic and thus $v(t)=1$; In the second and third regions the particle becomes nonrelativistic, with $v(t)=\langle p(t)\rangle/M_{N_1}$. During the radiation and matter dominating eras, the scale factor $R(t)=R_{eq}\sqrt{t/t_{eq}}$ and $R(t)=R_{eq}({t/t_{eq}})^{2/3}$, respectively. Here $t_{eq}$ and $R_{eq}$ are the quantities defined at the radiation-matter equality. The critical time $t_{nr}$ is detemirned by $\langle p(t)\rangle/M_{N_1}=1$ with $\langle p(t)\rangle=2.45 T$, which gives 
\begin{align}\label{}
t_{nr}=2.45^2\times 0.3g_{*}^{-1/2}M_{\rm Pl} /M_{N_1}^2\simeq 0.15\times 10^5\times\L\f{100}{g_{*}}\R^{1/2}\L\f{10\rm keV}{M_{N_1}}\R^{2}s,
\end{align}
which justifies the assumption $t_{nr}\ll t_{eq}=1.9\times10^{11}s$. Now, the nonrelativistic velocity is expressed as 
\begin{align}\label{}
v(t)=\sqrt{\f{t_{nr}}{t}}, ~(t_{nr}<t\leq t_{eq});\quad v(t)=\sqrt{\f{t_{nr}}{t_{eq}}}\L\f{t_{eq}}{t}\R^{2/3},~(t>t_{eq}).
\end{align}
With them, it is ready to calculate $\ld_{fs}$:
\begin{align}\label{FS}
\ld_{fs}={\cal S}^{-1/3}\f{\sqrt{t_{eq}}}{R_{eq}}\L2+\ln\L\f{t_{eq}}{{t_{nr}}}\R+3\R \sqrt {t_{nr}}.
\end{align}
Note that $t_{in}$ is at very early Universe, corresponding to temperature $m_{H_a}$ around the weak scale, thus there is a sizable entropy dilution factor ${\cal S}\equiv g_*(t_{in})/g_*(t_{0})\approx 33$, which redshifts the momentum of dark matter by a factor ${\cal S}^{-1/3}$ indicated above. It is not surprising that the above expression is the same to the one derived in Ref.~\cite{Merle:2013wta}, since the average momentums have the same scaling behave $\propto T$. The difference is manifested in the expression in $t_{nr}$.

\section{Real scalar as a FI$m$P}\label{SFImP}

In this appendix we consider a scalar FI$m$P $S$, which is supposed to be a real singlet scalar. Although it is not predicted by some well-motivated physics, it is still of great interest because of its even clearer way to show the main merits of FI$m$P, stability, common origins for mass and relic density. 

Different to the RHN case, now it can directly couple to the SM Higgs doublet via $\ld S^2|H|^2/2$, which gives mass to DM, $m_S=\sqrt{\ld /2}v$. This realizes the toy model considered in Ref.~\cite{Guo:2014bha}, where only singlets are introduced with some of them triggering EWSB and the lightest one being DM candidate. Here DM is not a WIMP but a FI$m$P, thus viable facing the strict direct detection bound. Concretely, 
in this single parameter model the freeze-in process is Higgs decaying into a pair of $S$, with decay width 
\begin{align}\label{}
\Gamma(h\ra SS)=\f{1}{32\pi}\f{\ld^2 v^2}{m_h}.
\end{align}
Then, the relic density, in terms of Eq.~(\ref{freezein}), is estimated to be
\begin{align}\label{}
\Omega h^2\simeq 0.12\times \L\f{\ld}{10^{-10.5}}\R^{5/2} \L\f{v/m_h}{2.0}\R^{3}\L \f{10^3}{g_{*}^S \sqrt{g_{*}^\rho}}\R.
\end{align}
But DM mass is proportional to $\ld^{1/2}$ instead of $\ld$, thus DM now is predicted to be $m_S\simeq 1.0$ MeV. So this FI$m$P is not warm DM-like (but it may elegantly account for a recent observation of DM self-interaction~\cite{Kang:2015aqa}). Varying the ratio $v/m_h$ in an extended Higgs sector, for example, the two Higgs doublet model with the extra doublet $H_2$ developing VEV around the GeV scale~\cite{Guo:2014bha}, the single term ${\ld_{12}}S^2{\rm Re}( H^\dagger H_2) $ may accommodate the scalar FI$m$P as a warm DM again.

\end{document}